\documentclass[12pt,preprint]{aastex}

\shorttitle{Multiple stellar populations in LMC clusters}
\shortauthors{Mackey et al.}

\begin{document}

\title{Multiple stellar populations in three rich Large Magellanic Cloud star clusters\altaffilmark{1}}


\author{A.D. Mackey\altaffilmark{2}, P. Broby Nielsen\altaffilmark{2}, A.M.N. Ferguson\altaffilmark{2}, J.C. Richardson\altaffilmark{2}}


\altaffiltext{1}{Based on observations made with the NASA/ESA Hubble Space 
Telescope, obtained at the Space Telescope Science Institute, which is operated 
by the Association of Universities for Research in Astronomy, Inc., under NASA 
contract NAS 5-26555.} 
\altaffiltext{2}{Institute for Astronomy, University of Edinburgh, Royal Observatory, 
Blackford Hill, Edinburgh, EH9 3HJ, UK}


\begin{abstract}
We present deep color-magnitude diagrams for three rich intermediate-age star 
clusters in the LMC, constructed from archival ACS F435W and F814W 
imaging. All three clusters exhibit clear evidence for peculiar main-sequence turn-offs.
NGC 1846 and 1806 each possess two distinct turn-off branches, while the turn-off for
NGC 1783 shows a much larger spread in color than can be explained by the photometric
uncertainties. We demonstrate that although all three clusters contain significant
populations of unresolved binary stars, these cannot be the underlying cause of
the observed turn-off morphologies. The simplest explanation is that each cluster
is composed of at least two different stellar populations with very similar metal
abundances but ages separated by up to $\sim 300$ Myr. The origin of these unusual
properties remains unidentified; however, the fact that at least three massive clusters 
containing multiple stellar populations are now known in the LMC suggests a potentially 
significant formation channel.
\end{abstract}



%% and \citet commands to identify citations. 

\section{Introduction}
It is a long-held notion that rich star clusters, including globular
clusters, are composed of single stellar populations -- that is, that each 
such object is made up of stars of a uniform age and chemical composition. 
New observations are challenging this accepted picture, however. 
It has recently been discovered that several of the most
massive Galactic globular clusters harbor multiple stellar populations
with a wide variety of unexpected characteristics. 
The most striking example is $\omega$ Centauri, which exhibits 
at least four different populations covering a wide spread in metal abundance
and age across the sub-giant branch (SGB) \citep{villanova:07}, as well
as a main sequence bifurcation which implies the
presence of a vastly helium-enriched population \citep{piotto:05}.
NGC 2808 possesses a triple main sequence split, suggesting two
helium-enriched populations \citep{piotto:07}; while NGC 1851
and NGC 6388 show clear splits in their SGBs, possibly indicative of 
populations with ages $\sim 1$ Gyr apart, or of sizeable intracluster
variations in chemical composition \citep{milone:08,piotto:08,salaris:08}.
In addition, several massive globular clusters in M31, including G1, 
apparently possess significant internal spreads in metal abundance 
\citep{meylan:01,fuentescarrera:08}.
Overall, these observed properties pose serious challenges 
for conventional models of globular cluster formation and evolution.

Rich intermediate-age star clusters in the Magellanic Clouds have the potential 
to open a new angle on this problem. These objects have masses comparable to many 
Galactic globulars (although are an order of magnitude less massive than
clusters like NGC 2808), but possess color-magnitude diagrams (CMDs) far more 
sensitive to internal dispersions in age, for example. We recently discovered 
that the massive LMC cluster NGC 1846 possesses a CMD exhibiting two clearly distinct
main-sequence turn-offs (MSTOs) \citep{mackey:07}, which we interpreted as belonging
to two stellar populations with similar metal abundance but ages $\approx 300$ 
Myr apart. Motivated by this discovery we have searched the HST/ACS archive for
additional observations of both NGC 1846 and other rich intermediate-age LMC
clusters. In this Letter we report on the first results from this search, presenting 
a new deep CMD for NGC 1846 as well as CMDs for two additional objects -- NGC 1783 
and 1806 -- that exhibit peculiar MSTOs. Similar results for these clusters have 
simultaneously and independently been obtained by another group \citep{goudfrooij:08}.

\section{Observations and data reduction}
ACS/WFC imaging of the three clusters was taken through the F435W and F814W filters 
as part of HST program GO 10595 (P.I. Goudfrooij), between 2005 September 29 -- 2006 
January 14. Each cluster was imaged three times per filter -- two long exposures of 
duration 340s each, and one short exposure of duration 90s in F435W and 8s in F814W. 

We used the ACS module of the {\sc dolphot} photometry software \citep{dolphin:00}
to photometer the images. {\sc dolphot} performs point-spread function (PSF) fitting 
using PSFs especially tailored to the ACS cameras, and works directly on 
flat-fielded images from the STScI archive relative to some deep reference image 
(we used the drizzled combination of the two long-exposure F814W images for each 
cluster). {\sc dolphot} accounts for the hot-pixel and cosmic-ray masking 
information provided with each flat-fielded image, fits the sky locally around each 
detected source, and provides magnitudes corrected for charge-transfer efficiency 
(CTE) effects on the calibrated VEGAMAG scale of \citet{sirianni:05}. 

For a given cluster we found that the narrowest CMD sequences were obtained 
by treating the four long-exposure images together and the two short-exposure images 
together in two separate {\sc dolphot} runs. We used the photometric quality 
information provided by {\sc dolphot} to clean the two resulting detection lists before 
combining them. In each list we selected objects of stellar type, with valid measurements 
on all input images, global sharpness parameter between $\pm 0.3$ in each filter, and 
``crowding'' parameter less than $0.5$ mag in each filter. We combined the
cleaned lists by selecting all long-exposure detections fainter than, and all short-exposure 
detections brighter than $0.5$ mag below the long-exposure saturation level at 
$m_{{\rm F435W}} \approx 19$. A small region of overlap between the lists was used to 
calculate and eliminate any small global offsets (typically\ $\la 0.02$ mag).

\section{Analysis}
\begin{figure}
\begin{center}
\epsscale{1.0}
\plotone{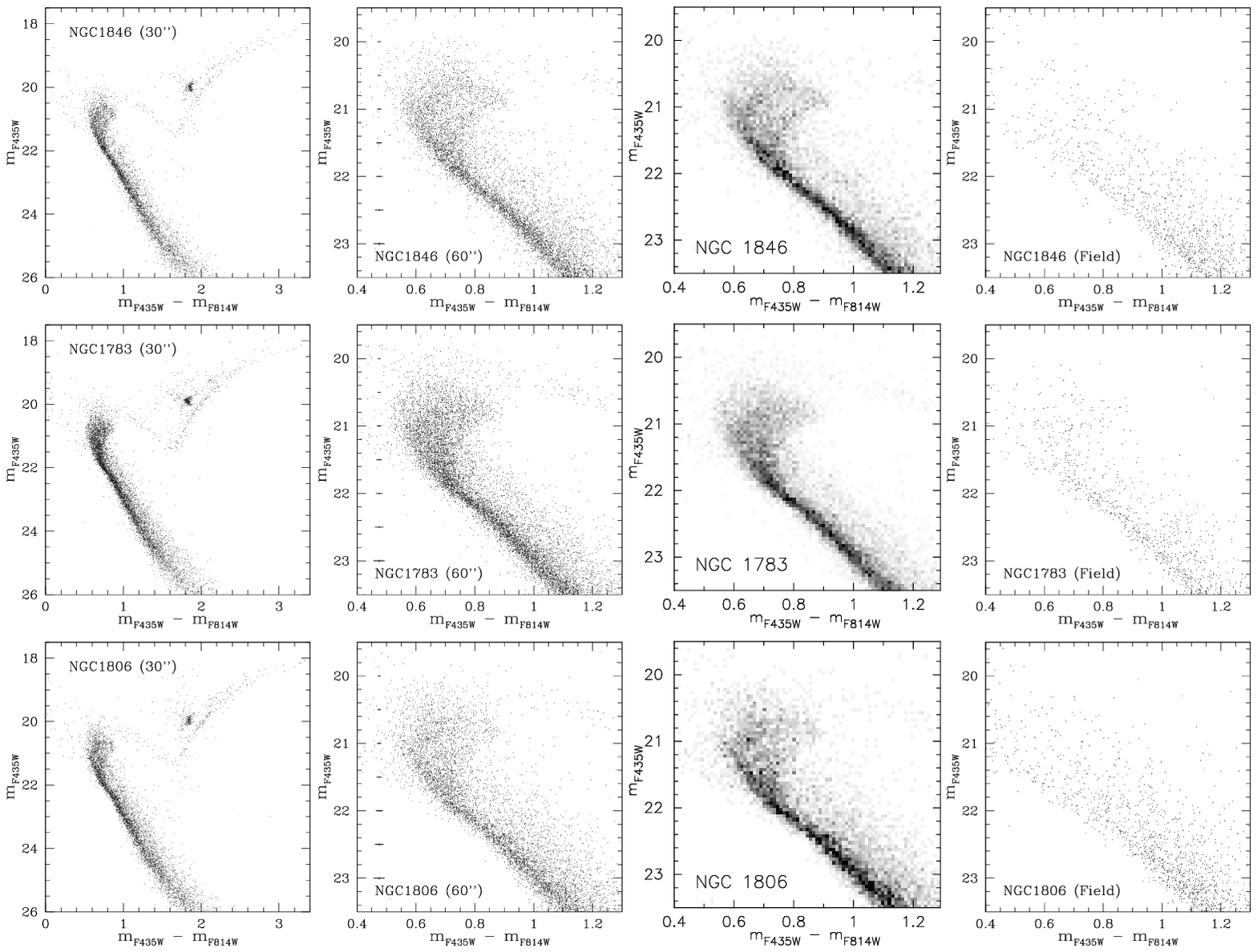}
\end{center}
\caption{ACS/WFC CMDs and Hess diagrams for NGC 1846, 1783, and 1806. Radial cuts
of $30\arcsec$ or $60\arcsec$ have been imposed as marked. In the second column the 
sizes of typical photometric uncertainties are indicated. Each Hess diagram has bin 
sizes of $0.01$ mag in color and $0.05$ mag in $m_{{\rm F435W}}$. The field CMDs
consist of stars taken as far as possible from each cluster center (typically beyond 
$\sim 120\arcsec$) over areas equivalent to those used for the central MSTO extractions.
\label{f:cmds}}
\end{figure}

Our CMDs for NGC 1846, 1783, and 1806 are presented in Fig.~\ref{f:cmds}. In these plots, 
we have isolated the cluster sequences from the surrounding low-level field populations by 
imposing radial cuts of $30\arcsec$ and $60\arcsec$ as marked. Full CMDs are shown in the 
left-hand column, while the second column shows close-ups of the cluster MSTOs and the
third column presents Hess diagrams of these MSTO regions. In the right-hand column are
CMDs constructed by taking stars as far as possible from each cluster center over areas
equivalent to those used for the central MSTO extractions. These outer CMDs likely still
contain a few cluster members, and are hence not completely pure field samples. Nonetheless,
they clearly demonstrate that the unusual CMD features described below are not the result 
of field star contamination.

All three clusters exhibit striking, unusual MSTOs. That for NGC 1846 possesses two clear
branches, as demonstrated previously by \citet{mackey:07} from shallower ACS observations
(S$/$N$\,\sim 150$ at the MSTO, compared with S$/$N$\,\sim 250$ here). The 
MSTO for NGC 1806 also possesses two clear branches, spaced similarly to those of NGC 1846. 
In contrast, the MSTO for NGC 1783 is not evidently split into branches; instead, the spread 
in color of the MSTO for this cluster is much larger than can be explained by the 
photometric uncertainties (marked in Fig.~\ref{f:cmds}; note also the narrowness of the 
main sequence below the turn-off). This may represent individual MSTO branches 
which are closely spaced and unresolved by the observations, or a smoothly spread distribution 
of stars. We find no evidence in the three clusters that stars belonging to different 
MSTO sections (e.g., upper/lower branches) possess different spatial distributions.
\citet{mucciarelli:07} previously noted a spread in color about the MSTO of NGC 
1783; however, their CMD was from shallower data such that the observed spread was 
consistent with that expected from the photometric uncertainties. 

Apart from the peculiar MSTO morphologies, the CMDs in Fig.~\ref{f:cmds} are as expected 
for intermediate-age clusters. The main sequences, sub-giant branches (SGBs) and red giant 
branches (RGBs) are narrow and the red clumps are well defined, implying that the MSTO 
structures are not due to significant line-of-sight depth or differential reddening 
in these clusters. The narrow CMD sequences further suggest minimal internal dispersions 
in $[$Fe$/$H$]$; however we note they place no constraints on the possibility of internal 
variations in other chemical abundances -- for example, CN, O, Na, or $[\alpha/$Fe$]$ -- 
as are observed for several of the peculiar massive Galactic globular clusters 
\citep[e.g.,][]{piotto:08}. 

Also evident on each cluster CMD is a clear spread of stars above and to the red of
the main sequence, implying a non-negligible population of unresolved binary systems
\citep[e.g.,][]{hurley:98}. To quantify this in a given cluster, we 
selected all stars within $30\arcsec$ of the center 
and in the range $22 \le m_{{\rm F435W}} \le 23.2$ (i.e., along the upper main sequence). 
We fit a ridgeline to the main sequence, and constructed a histogram of the spread in 
stellar color about this peak. The dispersion to the blue represents photometric 
uncertainties only; to the red, there is an additional component due to binaries.
Reflecting the blue-side dispersion to the red and subtracting this from the histogram 
thus allows the number of binary stars to be approximately determined. 
Because of the presence of photometric uncertainties, only binaries with mass ratios 
sufficiently large that they lie more than a few$\,\times\,0.01$ mag in color from 
the main sequence peak remain after this subtraction -- experiments with synthetic CMDs 
(see below) suggest a limiting mass ratio of $q \sim 0.5$. We measured binary fractions 
of $\approx 40\%$ for each of our three clusters, from $\sim1500$ stars per 
histogram. Unresolved binaries can strongly influence a cluster's MSTO morphology and 
it is therefore important to assess whether these significant inferred populations 
can explain the peculiar MSTOs. This is unlikely to be the case, since
inspection of Fig. \ref{f:cmds} reveals that each peculiar MSTO clearly branches from 
the {\it single star} main sequence rather than from the binary sequence.
We explore this question in more detail below with the aid of synthetic CMDs. 

The most straightforward interpretation of our CMDs is that each cluster possesses at 
least two stellar populations with very similar $[$Fe$/$H$]$ but differing ages. 
Since the possibility of variations in other chemical abundances is as-yet unconstrained, 
in what follows we adopt the simplest assumption of complete chemical homogeneity in each system.

To quantify the properties of each cluster's stellar populations in this scenario, we 
generated a grid of isochrones in the ACS/WFC (F435W, F814W) plane spanning certain 
ranges of age and $[$M$/$H$]$, and searched this grid to find the isochrone with a shape 
most closely matching that of the relevant CMD. We adopted recent isochrones from the 
Padova group \citep{marigo:08}, using their interactive web form 
({\it http://stev.oapd.inaf.it/cmd}) to generate a fine grid sampling an age 
range $1.0 \le \tau \le 3.0$ Gyr at intervals of $0.1$ Gyr, and a metal abundance range 
$0.0025 \le Z \le 0.0095$ at intervals of $\Delta Z = 0.0005$. These values are 
appropriate for NGC 1846 and 1783 which have $\tau \sim 1.5-2.0$ Gyr and $[$M$/$H$]\sim -0.4$ 
\citep[e.g.,][]{mackey:07,mucciarelli:07}, as well as for NGC 1806 which is poorly studied 
but possesses a very similar CMD to that of NGC 1846.

Full details of the isochrone fitting procedure are presented by \citet{mackey:07}. 
For a given cluster we first matched isochrones to the upper MSTO branch; for NGC 1783 we 
used the upper envelope of the MSTO with the aim of determining the maximum age spread 
allowed by the width of this feature. Having calculated the necessary shifts in color 
and magnitude required to overlay the best-fitting isochrone to the upper MSTO branch, 
we next applied our technique to the lower MSTO branch (the lower envelope of the MSTO 
for NGC 1783) with the added constraint that the best-fitting isochrone must have 
equivalent such shifts. This ensured a consistent solution for each cluster in the 
absence of any significant line-of-sight depth or differential reddening.

\begin{figure}
\begin{center}
\epsscale{1.0}
\plotone{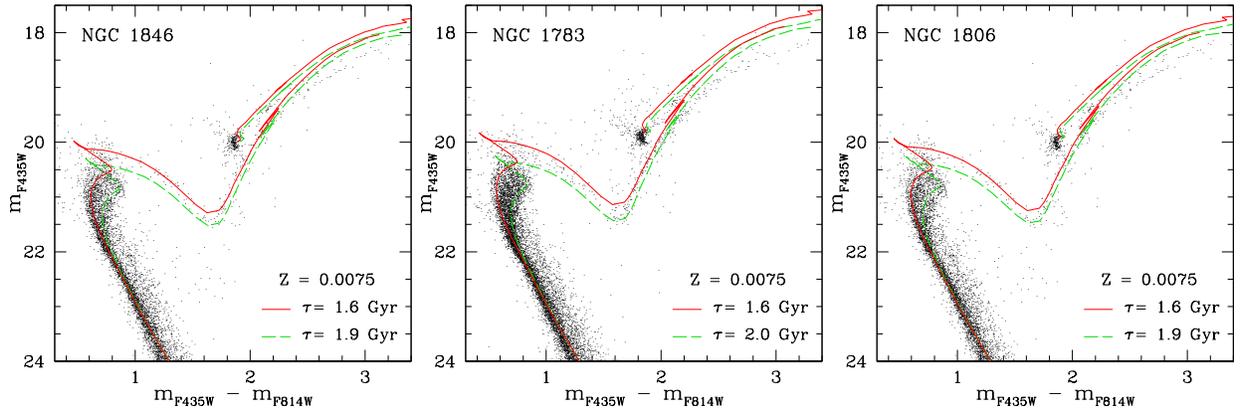}
\end{center}
\caption{Best-fitting isochrones from \citet{marigo:08} overlaid on the three cluster
CMDs.\label{f:isofit}}
\end{figure}

Our results are presented in Fig.~\ref{f:isofit}. In all three cases the highest 
quality solutions were for $Z=0.0075$ isochrones. This corresponds to 
$[$M$/$H$]\approx -0.40$, which, as noted above, matches previous spectroscopic and 
photometric estimates for both NGC 1846 and 1783. In NGC 1846 and 1806, 
the difference in age between the two MSTO branches is $\approx 300$ Myr, with the 
oldest branch representing a population of age $\sim 1.9$ Gyr and the youngest 
branch a population of age $\sim 1.6$ Gyr in both clusters. In NGC 1783, the upper 
and lower envelopes of the MSTO imply a maximum age spread of $\approx 400$ Myr, with
mean age $\sim 1.8$ Gyr. \citet{mucciarelli:07} previously found 
$\tau\approx 1.4\pm 0.2$ Gyr for this cluster using 
isochrones from the Pisa Evolutionary Library with mild overshooting.

Our best-fitting isochrones match the cluster main-sequences, MSTOs, SGBs, lower RGBs,
and red clumps well. The primary discrepancy between the isochrones and CMDs occurs
in each case on the upper RGB where the isochrones appear bluer than the observed stars.
This may represent some unidentified limitation in either the stellar models or our 
photometry. Selecting isochrones sufficiently metal-rich to closely trace the upper 
RGB sequences results in badly degraded fits to the other CMD regions, particularly
the main-sequences and MSTO shapes. Our fitting technique allows rough
uncertainties to be placed on our best values for age and metal abundance. Moving
to isochrones adjacent on the grid to the best models already leads to noticeably 
poorer fits -- hence we conservatively estimate random uncertainties of $\pm 0.001$ 
in $Z$ and $\pm 0.1$ Gyr in relative age.

We next constructed synthetic CMDs to investigate the role played by unresolved
binary stars around the MSTO regions, and in particular investigate whether these
objects might reproduce the observed MSTO structures. To do this we generated $5000$
stellar detections of which $2000$ ($40\%$) were unresolved binaries. These numbers 
approximately match those for the CMD of NGC 1846. We drew single star and 
primary masses (and hence luminosities) randomly from a uniform distribution along 
the $Z = 0.0075$, $\tau = 1.9$ Gyr isochrone, which best-fit the lower MSTO branch 
in NGC 1846. In the scenario where the MSTO structure is due to binaries, the lower 
MSTO branch must represent single stars and the upper branch the binary sequence. To 
produce the binary sequence we generated a secondary star for each primary by randomly 
selecting a mass ratio from a uniform distribution in the range $0.5 - 1.0$. 
This synthesis procedure is clearly simplistic -- more realistic would be to
generate stellar masses along the isochrone by randomly selecting from an appropriate 
mass function, as well as allowing a non-uniform distribution of mass ratios.
We emphasize, however, that our aim here is to observe the area on the CMD 
occupied by the binary sequence. Varying the generation algorithms presently represents 
an unnecessary level of complexity.

\begin{figure}
\begin{center}
\epsscale{0.7}
\plotone{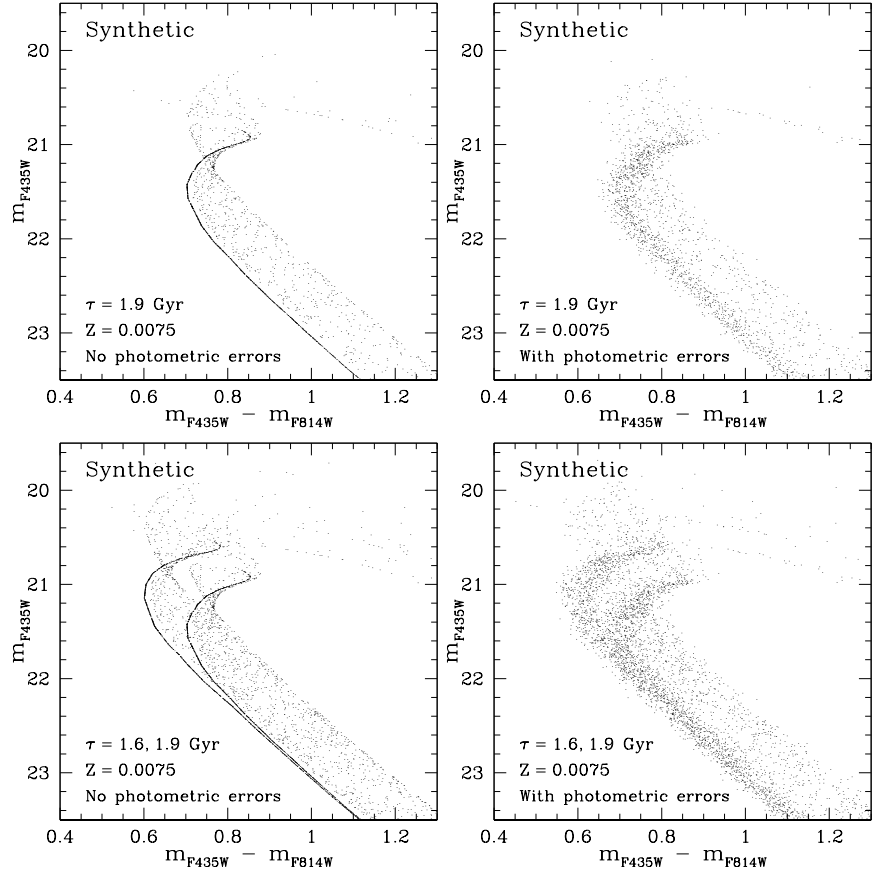}
\end{center}
\caption{Synthetic CMDs calculated as described in the text. A single 
stellar population cannot reproduce the MSTO structure of, e.g., 
NGC1846, even with a significant fraction of unresolved binaries; however 
including two separate populations results in a close match.\label{f:synth}}
\end{figure}

The upper row in Fig.~\ref{f:synth} shows our results. 
The left panel is the synthetic CMD generated as described above. In the right 
panel we added random photometric uncertainties to the F435W and F814W 
magnitudes of each star in this synthetic CMD by measuring the uncertainties 
determined by {\sc dolphot} for detected stars in NGC 1846 of equivalent 
brightness and color, and then generating errors from Gaussian distributions 
with these widths. These plots clearly demonstrate that unresolved binaries 
cannot alone reproduce the peculiar MSTO structures. 
Specifically, the binary sequence does not result in a distinct branch above and 
to the blue of the lower MSTO branch, as is observed for NGC 1846 (Fig. \ref{f:cmds}). 

We next altered our synthesis procedure to use both best-fitting isochrones 
for NGC 1846 (Fig.~\ref{f:synth}, lower panels).
This new CMD closely resembles the observed CMD, demonstrating that
two stellar populations in NGC 1846 (and by extension NGC 1806) are necessary 
to explain the observed MSTO structure. At least some of the spread of stars 
between the two branches in NGC 1846 must be due to unresolved binaries; however, 
it is intriguing that the two branches each appear to have a larger spread than 
those in the synthetic CMD, which are quite well defined even with the addition of 
photometric uncertainties. We do not assert here that this spread is intrinsic, 
merely that it is worthy of more detailed investigation. 

\section{Concluding remarks}
In this Letter we have demonstrated that three rich intermediate-age LMC star
clusters are comprised of multiple stellar populations. Several other LMC
objects with similar properties have previously been tentatively identified.
The MSTO in NGC 2173 ($\tau \sim 1.5$ Gyr) suggests a possible $\sim 300$ Myr 
internal age dispersion \citep{bertelli:03}, while NGC 1868 may exhibit a weak 
secondary MSTO \citep{santiago:02}. The young cluster NGC 2011 apparently 
possesses twin main sequences \citep{gouliermis:06}.

It is difficult to imagine how our three LMC clusters, each an order of magnitude 
less massive than the peculiar Galactic globulars, might have retained sufficient 
gas to undergo multiple widely-separated episodes of star formation. Previous
work has shown that the time-dependent potential of a forming star cluster may
allow a population of older field stars to be captured \citep[e.g.,][]{pflamm:07}; 
however, such models cannot convincingly account for the large fraction of captured 
stars ($\sim 50\%$) required for our clusters, nor the apparent homogeneity
in age and metal abundance of the older population within each cluster. 
A plausible alternative to these suggestions is that each of our three objects is 
the merger product of two or more clusters formed separately within a single giant 
molecular cloud \citep{mackey:07}. The LMC possesses numerous ``double'' star 
clusters (e.g., NGC 1850) and its low tidal field could facilitate such mergers. 
Large-scale realistic $N$-body modelling, along with measurements of the structures
and internal kinematics of the three present clusters should allow strong constraints 
to be placed on this possibility.

\acknowledgments
ADM and AMNF are supported by a Marie Curie Excellence Grant from the European 
Commission under contract MCEXT-CT-2005-025869. 
%We are grateful to the referee,
%Giampaolo Piotto, for constructive comments which helped improve this paper.


%% in parentheses, after the keyword, but they will not be verified.

{\it Facilities:} \facility{HST (ACS)}.


\clearpage

\end{document}